\documentstyle[preprint,eqsecnum,aps]{revtex}

\begin{document}
\draft
\preprint{HEP/123-qed}
\title{
Large Thermoelectric Power Factor in TiS$_{2}$ Crystal with Nearly Stoichiometric Composition
}
\author{H. Imai, Y. Shimakawa, and Y. Kubo}

\address{
Fundamental Research Laboratories, NEC Corporation, 34 Miyukigaoka,
 Tsukuba 305-8501, Japan
}

\date{\today}
\maketitle
\begin{abstract}
A TiS$_{2}$ crystal with a layered structure was found to have a large thermoelectric power factor.
 The in-plane power factor $S^{2}/ \rho$ at 300~K is 37.1~$\mu$W/K$^{2}$cm with resistivity
 ($\rho$) of 1.7 m$\Omega$cm and thermopower ($S$) of -251~$\mu$V/K,
 and this value is comparable to that of the best thermoelectric material, Bi$_{2}$Te$_{3}$ alloy.
  The electrical resistivity shows both metallic and highly anisotropic behaviors,
 suggesting that the electronic structure of this TiS$_{2}$ crystal has a quasi-two-dimensional nature.
 The large thermoelectric response can be ascribed to the large density of state just above the Fermi energy
 and inter-valley scattering.
 In spite of the large power factor, the figure of merit, $ZT$ of TiS$_{2}$ is 0.16 at 300~K, 
because of relatively large thermal conductivity, 68~mW/Kcm. 
However, most of this value comes from reducible lattice contribution. 
Thus, $ZT$ can be improved by reducing lattice thermal conductivity, e.g., 
by introducing a rattling unit into the inter-layer sites. 
\end{abstract}


\narrowtext
 Thermoelectric energy conversion, which transforms heat directly into electricity, 
has attracted much interest in recent years for possible applications to 
``environmentally friendly" electric-power generators and highly reliable, 
small-scale refrigerators used for electronic devices. 
The efficiency of a thermoelectric device is defined by its material properties through 
the dimension-less figure-of-merit, $ZT = S^{2}T/\rho ( \kappa_{e} +\kappa{_l})$, 
where $S$ is the Seebeck coefficient, $T$ is operating temperature, $\rho$ is resistivity, 
and $\kappa_{e}$ and $\kappa_{l}$ are carrier and lattice thermal conductivities, respectively. 
Although some thermoelectrics, such as Bi$_{2}$Te$_{3}$, are used in particular fields of application, 
their efficiencies, at the best $ZT$ of about 1, are not enough for wider use in commercial applications. 
Thus, development of materials with higher efficiency is one of the current main interests 
in research on thermoelectric materials.

A newly proposed concept, ``phonon-glass and electron-crystal"(PGEC), 
gives us clues to develop new materials with large $ZT$. 
PGEC has low thermal conductivity like glass and high electrical conductivity usually observed in crystals. 
Filled skutterudite antimonides~\cite{ref1,ref2,ref3} and Ge chraslates~\cite{ref4,Yb} are newly discovered PGEC. 
By introducing rattling atoms into parent electrically conductive cage (to make phonon glass state), 
we can produce large $ZT$ materials.  
Other examples of PGEC are superlattices,~\cite{ref5,ref6,ref7} 
in which the mean free path of heat-carrying phonons is considerably reduced 
because of their two-dimensional structure. 
In addition, a theoretical calculation predicted that 
a two-dimensional electronic structure enhances thermopower,~\cite{ref8} 
and such a structure is a great advantage of thermoelectrics. 
In the light of this PGEC concept, conducting compounds with naturally layered 
structure are candidate materials for thermoelectrics, 
if they have a large power factor, PF=$S^{2}/\rho$. 
They are expected to have intrinsically low $\kappa_{l}$, like supperlattices, and $\kappa_{l}$ 
can be reduced by introducing rattling units into inter-layer sites.

A layered-structure material TiS$_{2}$ exhibits metallic conductivity, 
thus it can be a good thermoelectric material. 
It has an anisotropic structure with a trigonal space group, P$\overline{3}$m, as shown in Fig. 1.~\cite{ref9} 
And it consists of infinite layers of edge-sharing TiS$_{6}$ octahedra. 
In the layers, TiS$_{6}$ octahedra are combined with each other tightly through strong covalent bonds, 
while each layer stacks weakly by van der Waals force. 
Transport properties of TiS$_{2}$ are strongly affected by 
its large off-stoichiometry.~\cite{ref10,ref11,ref12,ref13,ref14} 
Excess titanium atoms are intercalated into the van der Waals gap, 
and they introduce electrons into the host TiS$_{6}$ layers, 
giving rise to metallic conductivity with a carrier density of $10^{20}$ - $10^{21}$ cm$^{-3}$. 
Although the basic transport properties of TiS$_{2}$, and their dependence on chemical composition, have been extensively studied, the thermoelectric properties of TiS$_{2}$ have not been discussed, probably because metallic materials are not suitable for thermoelectrics. 

In this study, we investigate the electrical, thermal transport, and thermoelectric properties of TiS$_{2}$ 
single crystal with nearly stoichiometric composition, since it is expected to have a more two-dimensional nature than large-off-stoichiometric TiS$_{2}$ and, thus, good thermoelectric properties.

A single crystal (dimensions: $10 \times 5 \times 0.1~$ mm $^{3}$) 
was grown by a chemical-vapor-transport method using I$_{2}$ transport agent. 
In-plane resistivity ($\rho_{a}$) and out-of-plane resistivity ($\rho_{c}$) 
were measured by the van der Pauw and Montgomery methods, 
respectively, using a lock-in technique in the temperature range from 4 to 300~K. 
Thermopower in this temperature range was measured by the conventional constant-$\Delta T$ method. 
In-plane Hall resistivity up to 2~T was measured in the van der Pauw configuration 
by applying the magnetic field normal to the $c$-plane in the temperature range from 5 to 300~K. 
In-plane and out-of-plane thermal conductivities at room temperature 
were measured by the AC thermal diffusivity and the laser flush methods, respectively. 

Figure 2 shows the in-plane resistivity, $\rho_{a}$, out-of-plane resistivity, 
$\rho_{c}$, and anisotropy ratio, $\rho_{a}/\rho_{c}$. 
Both $\rho_{a}$ and $\rho_{c}$ exhibited metallic behaviors, and they almost depend on $T^{2}$ 
($\rho_{a} \propto T^{2.18}$, $\rho_{c} \propto T^{1.78}$). 
Figure 3 shows the in-plane carrier density, $n=1/R_{H} e$ and 
the inverse Hall mobility $1/\mu_{H} = \rho/R_{H}$ as a function of $T$. 
The carrier density at room temperature is $2.8 \times 10^{20}$ cm$^{-3}$ and almost independent of $T$. 
If the Ti$^{4+}$ state is assumed, this carrier density 
gives the amount of excess titanium to be 0.0001~atoms/f.u., 
which indicates that the chemical composition of this crystal is very close to the stoichiometry.~\cite{ref15} 
It is noted here that this carrier density in TiS$_{2}$ is still 
one order of magnitude larger than that in thermoelectric Bi$_{2}$Te$_{3}$ with optimum carrier density. 

The resistivity-anisotropy ratios are large: 750 at 300~K and 1500 at 5~K. 
These ratios suggest that doped electron carriers are confined within the TiS$_{6}$ layers, 
thus, the electronic structure of TiS$_{2}$ is quasi two-dimensional. 
In-plane Hall mobility strongly depends on $T$ ($1/\mu_{H} \propto T^{2.14}$) 
and decreases with increasing $T$ from 150~cm$^{2}$/Vs at 5~K to 15~cm$^{2}$/Vs at 300~K. 
The significant increase in $\rho$ with increasing $T$ 
arises from the increase in scattering rate ($\propto 1/\mu_{H}$). 
The metal-like temperature dependence of $\rho_{c}$ in the order of about 1 $\Omega$cm is quite unusual. 
Although we cannot explain the dependence of $\rho_{c}$ on $T$ satisfactorily, 
the random network of marginally metallic paths along the $c$-axis between the metallic layers can give rise to it.

Figure 4 shows the dependence of in-plane thermopower, $S$, on temperature. 
The thermopower shows a broad peak near 30~K due to the phonon-drag effect 
resulting from the strong electron-phonon interaction.~\cite{ref16} 
Above 50~K, $S$ linearly depends on $T$, and reaches a large negative value, -251~$\mu$V/K at 300~K. 
Since the same order of large $S$ (about 200~$\mu$V/K) has been measured in Bi$_{2}$Te$_{3}$-Sb$_{2}$Te$_{3}$ 
system with carrier density of $3 \times 10^{19}$~cm$^{-3}$, 
$S$ of -251~$\mu$V/K is a surprisingly large value for such a high-carrier-density material.

 To understand the above described transport properties, 
a characteristic electronic structure of TiS$_{2}$ should be considered. 
An electronic band structure calculation revealed that a primary Ti $3d$ band crosses the Fermi level, 
making multi-valley structures with six small electron pockets 
around the $L$-point in the hexagonal Brilluion zone. 
In such a multi-valley electronic structure, inter-valley scattering 
significantly contributes to transport properties at higher temperatures, 
since it requires phonons with a large wave vector (almost half of the Brillouion zone).~\cite{ref10,ref17} 
Taking into account the inter-valley scattering, 
we can explain the observed nearly-$T^{2}$ dependence of $\rho_{a}$ as discussed in Ref. 11. 
If we assume simple intra- and inter-valley scattering with acoustic phonons, 
the inverse of total relaxation time is given by $1/\tau_{total} = 1/\tau_{intra}+1/\tau_{inter}$. 
Here, $1/\tau_{intra}$ linearly changes with $T$, 
while $1/\tau_{inter}$ is roughly proportional to $1/(\exp (\Theta_{D}/T)-1)$, 
where $\Theta_{D}$ is the Debye temperature.~\cite{ref10} 
Thus, the temperature dependence of total resistivity 
depends on the ratio of inter- and intra-valley scatterings, 
and resistivity nearly depends on $T^{2}$ 
when the intra- and inter-valley equally contribute to the resistivity.~\cite{ref10}  

Concerning large $S$ values, the density-of-state above the Fermi level 
originated from the conduction band plays an important role. 
Due to the large density-of-state just above the Fermi level, 
each electron pocket could contribute to large $S$.~\cite{ref18,ref19,ref20,ref21} 
The phonon-mediated inter-valley scattering also enhances $S$ 
by giving additional scattering channels to the conduction electrons in electron pockets; 
that is, it increases the entropy of carriers. 
Besides, the strong electron-phonon coupling contributes to enhancing $S$ by assisting the inter-valley scattering.

The power factor, $S^{2}/\rho$ of the TiS$_{2}$ crystal is large as shown in the inset of Fig. 4.  
The PF has a maximum at 52.3~$\mu$W/K$^{2}$cm near 30~K, where the phonon-drag effect occurs. 
Above 50~K, the PF is almost constant, because $S$ and $\rho$ depend on $T$ and $T^{2}$, 
respectively, and its value at 300~K is 37.1~$\mu$W/K$^{2}$cm. 
This PF is comparable to that of the best thermoelectric material, Bi$_{2}$Te$_{3}$. 

The measured in-plane and out-of-plane thermal conductivities 
at room temperature are 67.8~mW/Kcm and 42.1~mW/Kcm, respectively. 
The anisotropy ratio of the thermal conductivity is much smaller than that of the electrical conductivity. 
Because of these large thermal conductivities, 
the thermoelectric figure of merit is small, 0.16, at 300~K. 
The thermal conductivity by carriers is estimated 
by using the Wiedemann-Franz law to be 4.3~mW/Kcm (in-plane) and 0.006~mW/Kcm (out-of-plane) at 300~K. 
The significant difference between the measured $\kappa$ and the estimated $\kappa_{e}$ 
implies that most of the thermal conductivity comes from the lattice component. 
Thus, the anisotropy in $\kappa$ corresponds to the anisotropy in $\kappa_{l}$, 
which reflects the anisotropic crystal structure of TiS$_{2}$: 
that is, the TiS$_{6}$ octahedral layers provides a highly conductive phonon path in the plane, 
while the van der Waals gap reduces the phonon vibrations in the $c$-direction.

It is interesting to compare the thermoelectric and transport properties of TiS$_{2}$ 
with those of other thermoelectric materials.~\cite{ref4,Yb,ref22,rev1,rev2} 
As listed in Table 1, PF of TiS$_{2}$ at 300~K is 37.1~$\mu$W/K$^{2}$cm, 
which is comparable to that of the best thermoelectric Bi$_{2}$Te$_{3}$ compound,~\cite{ref22} 
though the carrier density of TiS$_{2}$ is one order of magnitude larger than that of Bi$_{2}$Te$_{3}$. 
Despite the large PF, the figure of merit $ZT$ of TiS$_{2}$ is small, 
0.16 at 300~K, because $\kappa$, 67.8~mW/Kcm, is about four times larger than that of Bi$_{2}$Te$_{3}$. 
However, the large $\kappa$ of TiS$_{2}$ is mainly contributed by the lattice component, 
and this is in sharp contrast with $\kappa$ of Bi$_{2}$(Te,Se)$_{3}$, 
where $\kappa_{e}$ is dominant. 
Hence, as found in some filled skutterudite antimonides,~\cite{ref1} 
the thermoelectric figure of merit of TiS$_{2}$ can be improved
 if $\kappa_{l}$ is reduced by introducing rattling units. 
In CoSb$_{3}$, by introducing heavy rattling lanthanide atoms 
into empty sites of the CoSb$_{3}$ cage, $\kappa_{l}$ has been reduced by one-eighth, 
so $ZT$ reaches more than unity at high temperature. 
In TiS$_{2}$, various guest species, such as alkali metals, 3$d$-transition metals, 
and organic molecules, can be intercalated into the van der Waals gaps.~\cite{rev4}
These intercalated units weakly couple with conductive TiS$_{6}$ layers 
and can reduce the lattice thermal conductivity by rattling or by introducing random scattering centers.
Hence, the $ZT$ of TiS$_{2}$ can be increased.
Finally, we note that PF of TiS$_{2}$ is large over a wide temperature range from 40~K to room temperature. 
This feature is a great advantage in applications such as very low-temperature coolers at, 
in particular, below liquid-N$_{2}$ temperatures.

In conclusion, by investigating transport and thermoelectric properties of ``old material" TiS$_{2}$ from the view point of a ``new concept", we found that TiS$_{2}$ crystal with a nearly stoichiometric composition has a large power factor, 37.1$-$52.3~$\mu$W/K$^{2}$cm, 
which is comparable to that of the best thermoelectric, Bi$_{2}$Te$_{3}$, over the wide temperature range.
The characteristic electronic state and scattering mechanism of TiS$_{2}$ due to the two-dimensional electronic state give the significantly large power factor.
Although its thermal conductivity is large compared to other thermoelectrics, most of it comes from the
reducible lattice component. Thus two-dimensional metallic TiS$_{2}$ can be a good thermoelectric, when the PGEC concept is applied. 
\begin{figure}
\caption{
Crystal structure of TiS$_{2}$ (Space Group P$\overline{3}$m1) and TiS$_{6}$ octahedral unit.
}
\label{Fig1}
\end{figure}

\begin{figure}
\caption{
Temperature dependence of in-plane resistivity, $\rho_{a}$, and  out-of-plane resistivity, 
$\rho_{c}$. The inset shows the anisotropy ratio, $\rho_{c}$ /$\rho_{a}$.
}
\label{Fig2}
\end{figure}

\begin{figure}
\caption{
Temperature dependence of in-plane carrier density and inverse Hall mobility.
}
\label{Fig3}
\end{figure}

\begin{figure}
\caption{
Temperature dependence of in-plane thermopower. The inset shows the thermoelectric power factor.
}
\label{Fig4}
\end{figure}


\mediumtext


\begin{table}
\caption{Thermoelectric and transport properties of single crystalline 
TiS$_{2}$, Bi$_{2}$Te$_{3}$-Bi$_{2}$Se$_{3}$, Ge chraslates,
and Co-based skutterudeite antimonides at 300~K. 
Note that La(FeCo)Sb$_{3}$ and Yb$_{0.02}$Co$_{4}$Sb$_{12}$ have $ZT$ of about 1 at high temperatures.
} 
\begin{tabular}{ccccccccccc}
Sample&$\rho$~&$S$&$\kappa$~($\kappa_{e}$)&$ZT$&$S^{2}/\rho$&$n$&$\mu$&Ref.\\
~&(m$\Omega$cm)&($\mu$V/K)&~(mW/cmK)&~&($\mu$W/K$^{2}$cm)&~(10$^{20}$cm$^{-3}$)&(cm$^{2}$/Vs)&~\\
\tableline
TiS$_{2}$&1.7&-251&67.8~(4.3)&0.16&37.1&2.8&15&this work\\
TiS$_{2}$~(30~K)&0.15&-87.2&-&-&52.3&3.0&133&this work\\
Bi$_{2}$Te$_{2.85}$Se$_{0.15}$&1.1&-223&15.9(6.7)&0.85&44.8&0.4&250&\cite{ref22,rev1}\\
Bi$_{1.65}$Te$_{3}$&1&-240&20.2~(7.4)&0.86&57.6&0.23&212&\cite{ref22}\\
Sr$_{8}$Ga$_{16}$Ge$_{30}$&10.5&-320&8~(0.7)&0.34&9.75&0.001&2200&\cite{ref4,rev2}\\
CoSb$_{3}$&14.2&-452&102~(0.5)&0.04&14.4&0.004&101&\cite{rev3}\\
La(FeCo)Sb$_{3}$&1.5&100&12~(4.9)&0.16&6.66&50&30&\cite{ref1}\\
Yb$_{0.02}$Co$_{4}$Sb$_{12}$&0.6&-150&45~(12.3)&0.3&37.5&-&-&\cite{Yb}\\
\end{tabular}
\label{table1}
\end{table}


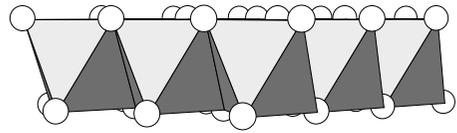
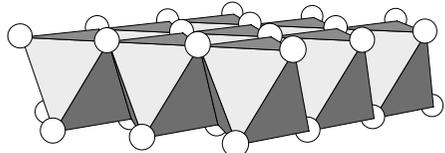
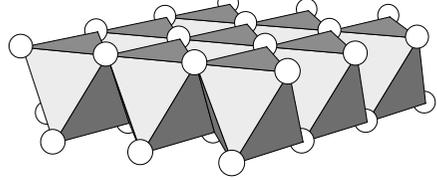
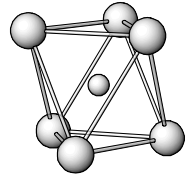
\end{document}